\documentclass[fleqn,10pt]{wlscirep}
\usepackage[utf8]{inputenc}
\usepackage[T1]{fontenc}
\usepackage{graphicx}%
\usepackage{multirow}%
\usepackage{amsmath,amssymb,amsfonts}%
\usepackage{amsthm}%
\usepackage{mathrsfs}%
\usepackage[title]{appendix}%
\usepackage{xcolor}%
\usepackage{textcomp}%
\usepackage{manyfoot}%
\usepackage{booktabs}%
\usepackage{algorithm}%
\usepackage{algorithmicx}%
\usepackage{algpseudocode}%
\usepackage{listings}%

\title{Pattern change of precipitation extremes in Bear Island}
\author[1,2,3,+]{Arnob Ray}
\author[4,5,+]{Tanujit Chakraborty}
\author[6]{Athulya Radhakrishnan}
\author[7]{Chittaranjan Hens}
\author[8]{Syamal K. Dana}
\author[1,*]{Dibakar Ghosh}
\author[6]{Nuncio Murukesh}
\affil[1]{Physics and Applied Mathematics Unit, Indian Statistical Institute, Kolkata, 700108, India}
\affil[2]{Roux Institute, Northeastern University, Portland, 04101, USA}
\affil[3]{Institute for Experiential AI, Northeastern University, Boston, 02115, USA}
\affil[4]{Department of Science and Engineering, Sorbonne University, Abu Dhabi, United Arab Emirates}
\affil[5]{Sorbonne Center for Artificial Intelligence, Sorbonne University, Paris, 75005, France}
\affil[6]{Ocean Atmospheric Studies, National Centre for Antarctic and Ocean Research, Vasco da Gama, 403804, India}
\affil[7]{Center for Computational Natural Science and Bioinformatics, International Institute of Informational Technology, Hyderabad, 500032, India}
\affil[8]{Division of Dynamics, Lodz University of Technology, 90-924 Lodz, postcode, Poland}

\affil[*]{diba.ghosh@gmail.com}
\affil[+]{these authors contributed equally to this work}

\keywords{Precipitation extremes, Arctic region, Extreme value theory, Generalized extreme value distribution, return levels, wavelet transform coherence}

\begin{abstract}
Extreme precipitation in the Arctic region plays a crucial role in global weather and climate patterns. Bear Island (Bj{\o}rn{\o}ya) is located in the Norwegian Svalbard archipelago, which is, therefore, selected for our study on extreme precipitation. The island occupies a unique geographic position at the intersection of the high and low Arctic, characterized by a flat and lake-filled northern region contrasting with mountainous terrain along its southern shores. Its maritime-polar climate is influenced by North Atlantic currents, resulting in relatively mild winter temperatures. An increase in precipitation level in Bear Island is a significant concern linked to climate change and has global implications. We have collected the amount of daily precipitation as well as daily maximum temperatures from the meteorological station of Bj{\o}rn{\o}ya located on the island, operated by the Norwegian Centre for Climate Services in Svalbard for a period spanning from January 1, 1960 to December 31, 2021. We observe that the trend of yearly mean 
precipitation during this period linearly increases. Also, the number of extreme precipitation events is also increasing. Motivating from this observation, we analyze the recorded data to investigate the changing pattern of precipitation extremes over the climate scales. We employ the generalized extreme value distribution to model yearly and seasonal maxima of daily precipitation amount and determine the return levels and return period of precipitation extremes. 
We compare the variability of precipitation extremes between the two time periods: (i) $1960$-$1990$ and (ii) $1991$-$2021$. Our analysis reveals an increase in the frequency of precipitation extremes occurrences between 1991 and 2021.  
 A  relationship between temperature and precipitation in the area is discerned using wavelet transform coherence. Our findings establish a better understanding of precipitation extremes in Bear Island from a statistical viewpoint, with an observation of seasonal and yearly variability, especially, during the period of the last 31 years. It sheds light on the changing climate in the Arctic region in the time span of $62$ years.
\end{abstract}
\begin{document}

\flushbottom
\maketitle
%
%
\thispagestyle{empty}


\section{\label{sec:intro} Introduction}
Global climate change increases the occurrence of precipitation extremes \cite{albeverio2006extreme, portner2022climate} and the change in patterns significantly affects the livelihood, agricultural productivity, and water availability of an area. Excessive rainfall brings detrimental consequences such as floods, crop damage, and disease outbreaks \cite{kangalawe2017climate, huq2004mainstreaming}.  Such changes are, particularly, evident in the Arctic, where extreme precipitation affects several components of the Arctic system, including river discharge into the Arctic Ocean, poleward moisture transport,  and amplified ice surface melt \cite{vihma2016atmospheric, walsh2020extreme, bintanja2020strong}. Capturing the precipitation trends in the Arctic provides valuable insights into the impacts of climate change on a global scale \cite{przybylak2003climate}. The melting of Arctic glaciers and ice caps, influenced by precipitation changes, affects ocean circulation, and weather patterns, and contributes to rising sea levels, which have implications for low-lying islands and coastal areas \cite{box2019key, landrum2020extremes}.

\par
\begin{figure}
		\centerline{
	\includegraphics[scale=0.4]{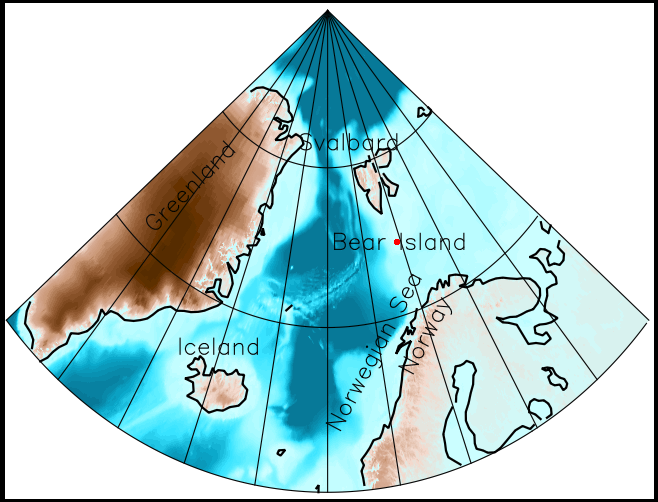}
 }
	\caption{Location of Bear Island (74.4522° N, 19.1152° E) is denoted by a red solid circle.}
	\label{fig1}
\end{figure}

 Our present concern is precipitation extremes in Bear Island (Bj{\o}rn{\o}ya), which is located in the western Barents Sea (marked with a red circle in Fig.\ \ref{fig1}), and it holds significant importance in the region between the high and low Arctic. The island serves as an ideal location for studies of extreme precipitation that may contribute to climate change and possesses various environmental attributes. Recent dendrochronological records of Bear Island have revealed the vulnerability of tundra plants to climate changes \cite{owczarek2020post}. While several studies have examined models for extreme precipitation in various geographic locations worldwide \cite{ghosh2012lack, santos2016estimating, ngailo2016modelling, yozgatligil2018extreme, tabari2021extreme}, the exploration of extreme precipitation patterns in the arctic region remains relatively limited. Few locations in the Arctic become the places for case studies of precipitation extremes in recent studies \cite{murukeshcharacteristics, nuncio2023hails}. Given the catastrophic nature of extreme weather events, in general, and their impact on the economy and infrastructure, a comprehensive understanding of such events is crucial for early detection and preparedness. From this perspective, we explore the intense precipitation events occurring in Bear Island from recorded data and attempt to analyze them using the extreme value theory (EVT) \cite{gnedenko1943distribution, kotz2000extreme, bali2003generalized, ghil2011extreme, engeland2004practical, poon2004extreme, mcneil1997estimating}.
maximum values observed in distinct blocks or intervals of the dataset.

EVT provides a framework for statistical analysis of extremes using either of the two basic approaches, the block maxima (BM) approach and the peaks-over threshold (POT) approach \cite{coles2001, lucarini2016extremes}. In BM approach, the maxima (or minima) observed from arbitrarily selected identical non-overlapping blocks in a time series are extracted as a data set for analysis. In the POT method, the event that exceeds a specified threshold is only considered as an extracted data set of extreme events. 
The distribution of events is typically modeled by the Generalized Extreme Value (GEV) distribution using the BM method, which is emphasized here for our statistical analysis. 

\par The primary focus of this study is to characterize the pattern of precipitation extremes statistically in Bear Island (Bj{\o}rn{\o}ya)  using recorded observations on daily precipitation data during the years from $1960$ to $2021$. We initiate our analysis with collected data of the annual maximum precipitation events and estimate the return level of precipitation extremes. A difference in the distributions of precipitation extremes is noticed between the two time intervals of 31 years: $[1960, 1990]$ and $[1991, 2021]$.  
The seasonal maximum precipitation events for the whole period [1960, 2021] as well as two separated climate time scales have been analyzed. By comparing two climate time scales, we conclude that the amount of precipitation extremes is higher in the past $31$ years. Even, these are going to increase in the future also. 
Furthermore, we explore the relationship between precipitation extremes and yearly mean maximum temperature anomaly using wavelet coherence \cite{torrence1998practical}.
\par In Sec.\ \ref{sec:data} of the paper, we describe about the location of Bear Island and the motivation of our work. In  Sec.\ \ref{sec:Methodology}, we elaborate on the model of Generalised extreme value distribution that is used for characterized precipitation extremes.   We exhibit our findings that the pattern of precipitation extremes is changing from one climate timescale to another in  Sec.\ \ref{sec:Resut}. Finally, we end with
the conclusions in Sec.\ \ref{sec:discussion}.

\section{\label{sec:data} Location and motivation}
\par Bear Island (Bj{\o}rn{\o}ya) is located within the high Arctic Svalbard archipelago, situated in the western Barents Sea between mainland Norway and the North Pole as shown in Fig.\ \ref{fig1}. As the southernmost island in the archipelago, Bear Island spans approximately 20 km in length and nearly 15 km in width. Its unique geography consists of flat, slightly hilly terrain in the northern and western regions, adorned with numerous lakes, while the remainder of the island is characterized by mountainous landscapes \cite{worslex2001geological}. Bear Island experiences a maritime-polar climate, influenced by the North Atlantic current \cite{owczarek2020post}. This climatic setting, combined with its isolated location, makes Bear Island an ideal location for studying extreme precipitation patterns and their possible effects within an Arctic context. 
\par
\begin{figure}
		\centerline{
	\includegraphics[scale=0.45]{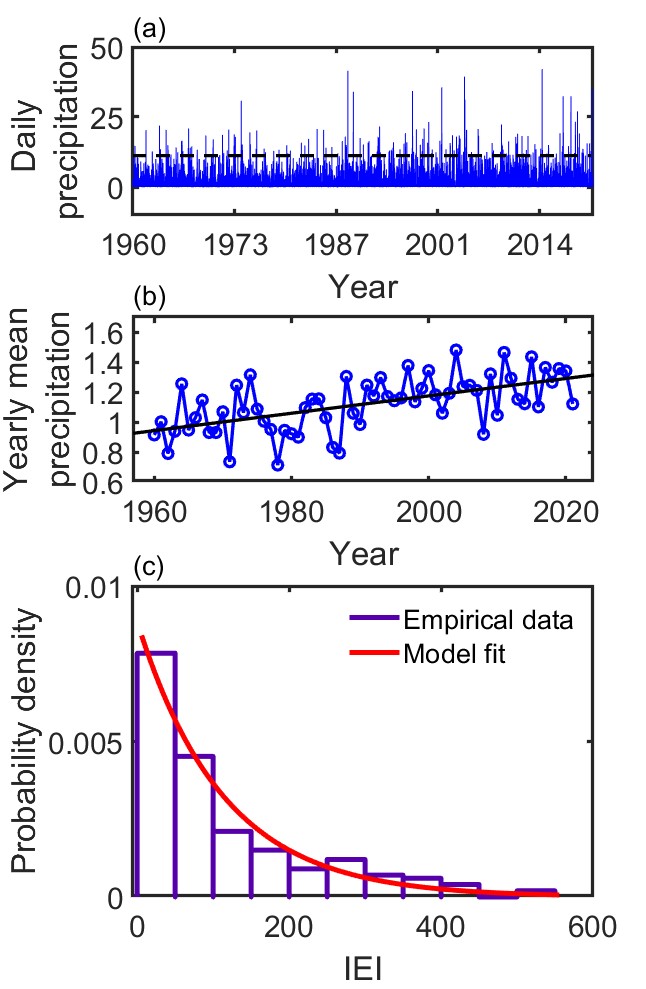}
 }
	\caption{(a) Temporal records of daily precipitation over the years $[1960, 2021]$. The $99$ percentile of the dataset consisting of daily precipitation over the period of $62$ years is $11.2$ which is considered as the threshold (horizontal dashed black line) for defining extreme precipitation events. (b) The variation of annual mean precipitation for $62$ years exhibits a linearly increasing trend fitted with a straight line (black) with a slope of $0.005794^o$C/year. (c) A histogram plot of inter-event intervals of extreme precipitation events is fitted with an exponential probability distribution function (red) following an exponential decay. Here, the estimated parameter ($\lambda$) of the exponential distribution is 0.0089.}
	\label{fig2}
\end{figure}

 We have collected daily precipitation as well as daily maximum temperature from the meteorological station of Bj{\o}rn{\o}ya (station no. SN99710) operated by the Norwegian Centre for Climate Services in Svalbard. The data span from January 1, 1960, to December 31, 2021, resulting in a dataset of 22,646 data points. The number of missing values in the dataset is minimal, with only 21 missing data points (0.092\%). Due to a negligible number of missing values, we have excluded them from the data set during the data preparation stage.  The temporal record of daily precipitation over the years $[1960, 2021]$ is shown in Fig.\ \ref{fig2}(a). 

A threshold-based technique is utilized to characterize extreme precipitation events. Extreme events are described as those that exceed the $99$th percentile of the considered datasets \cite{schar2016percentile, mcphillips2018defining, mishra2020routes}. The $99$ percentile of the whole dataset is drawn as a threshold (dashed black line) in Fig.\ \ref{fig2}(a). Clearly, we observe the higher frequency of extreme precipitation events occurring in the last few decades. Also, annual mean precipitation for the $62$ years exhibits a linearly increasing trend fitted with a straight line (black) with a slope of $0.005794$ as shown in Fig.\ \ref{fig2}(b). In order to establish the presence of a trend in the dataset, we additionally perform the Mann-Kendall (MK) test \cite{gilbert1987statistical} (described in Appendix\ \ref{sec:appendix}). For our case, $p$-value = $0.000006032 \; (< 0.05)$. Consequently, as the years go by, the mean amplitude of the precipitation gradually increases. This also concludes that extreme precipitation events are therefore anticipated to happen more frequently. We also discuss the return time statistics of precipitation extremes. For this, the probability density of the empirical data of inter-arrival times for extreme precipitation that occurs between $1960$ and $2021$ is now being plotted in Fig.\ \ref{fig2}(c). The time difference between two consecutive extreme events is considered as inter-arrival time or inter-event intervals \cite{bunde2005long}. Here, we consider one extreme event if heavy precipitation occurs two days or three days consecutively to avoid the clustering effect. The first day of those two or three days is taken into account for determining the day on which a heavy precipitation event occurred. This density plot is well-fitted by the probability density of exponential distribution which is given by 

\begin{equation}\label{eq.3}
\begin{array}{lcl} f(x;\lambda)=
\begin{cases} 
\lambda e^{-\lambda x}, &  ~x \ge 0,  \\
0, & ~x <0,
\end{cases}
\end{array}
\end{equation}
where $\lambda$ is the shape parameter. We determine the coefficient of variation (CV), which is derived by the ratio of standard deviation to the mean of the data. Theoretically, CV equals $1$ for the case of the exponential distribution \cite{heeger2000poisson}.  We calculate numerically CV = $1.031$ for our dataset consisting of inter-event intervals. Since, we are familiar that if events occur independently, the probability distribution of their time differences will follow an exponential distribution. Therefore, in our case, the precipitation extremes happen independently. So, our initial observation of time domain data and basic statistical analysis on daily precipitation and recognition of the existence of extreme precipitation events, demands a more rigorous statistical analysis using extreme value theory which we attempt here mainly using the BM method for characterization of precipitation extremes. We address the problem of the variability of precipitation extremes by changing the climatological time scale in the specific location. Also, we give an insight of the predictability of precipitation extremes in the future through our study.

\section{\label{sec:Methodology} Methodology: Block maxima approach}
A brief overview of the BM approach is presented along with the GEV approach as an asymmetric statistical distribution for extreme values \cite{coles2001}. The BM method is a technique commonly employed to identify extreme values within a dataset. In this method, the long time series of events, consisting of $n$ data points, is divided into a large number of identical blocks, denoted by $b >> 1$, each containing a smaller number of data points, denoted by $k >> 1$, such that $n = bk$. Within each block, the maximum value is determined, representing an extreme value or event. By applying the BM method for our case study, we collect the maximum amount of precipitation from each year, resulting in a set of annual maximum precipitation data series. For our specific case, we define the blocks as seasons to understand seasonal fluctuations in the data. The extreme precipitations collected using the BM method are assumed to follow the GEV distribution, which characterizes the probability distribution of extreme precipitation.

\par To explain the GEV technique, let us consider an independent and identically distributed (i.i.d.) random variables $Y_1,$ $Y_2,$ $Y_3,$ $\dots,$ $Y_n$ with a cumulative distribution function $G(y)~(=P(Y_i\le y))$ in common. Let, $M_n=max\{Y_1,~Y_2,~ Y_3,\cdots,~Y_n\}$ be another random variable. Using the Fisher–Tippett–Gnedenko theorem \cite{gnedenko1943distribution}, we write
$$ P\Big(\dfrac{M_n-b_n}{a_n}\le y\Big)\rightarrow G(y) \; \; \text{as} \; \; n\rightarrow\infty,$$
where \{$a_n>0$\} and \{$b_n$\} are normalizing constants. It is known that $G(y)$ belongs to the family of GEV distribution which can be mathematically described as,

\begin{equation}
\label{eq.1} 
G(y; \mu,\sigma,\xi)=
	\begin{cases}
	\exp{\bigg[-\bigg(1+\xi \dfrac{y-\mu}{\sigma}\bigg)^{-\frac{1}{\xi}}\bigg]}; &\; \text{if} \; \xi \neq 0 \; \text{and} \bigg(1+\xi \dfrac{y-\mu}{\sigma}\bigg)>0,\\
	\exp\bigg[-\exp \bigg\{-\dfrac{y-\mu}{\sigma}\bigg\}\bigg];& \; \text{if} \; \xi =0 \; \text{and} -\infty < y < \infty.
	\end{cases}
\end{equation}

Here $\mu~(-\infty<\mu<\infty)$ is the location parameter, $\sigma ~(\sigma>0)$ is the scale parameter and $\xi~(-\infty<\xi<\infty)$ is the shape parameter. GEV has three types of domain of attraction depending on the sign of $\xi$. If $\xi=0$, then $G(y)$ belongs to the Gumbel family of distribution that implies a light tail or exponential tail of the distribution. When $\xi>0$ then $G(y)$ exists in the Fr{\'e}chet family of distribution that indicates the heavy or fat tail of the distribution. For $\xi<0$, $G(y)$ follows the Weibull family of distribution that demonstrates a short tail or upper finite end point of the distribution \cite{chakraborty2022searching}.
\begin{figure*}
		\centerline{
	\includegraphics[scale=0.55]{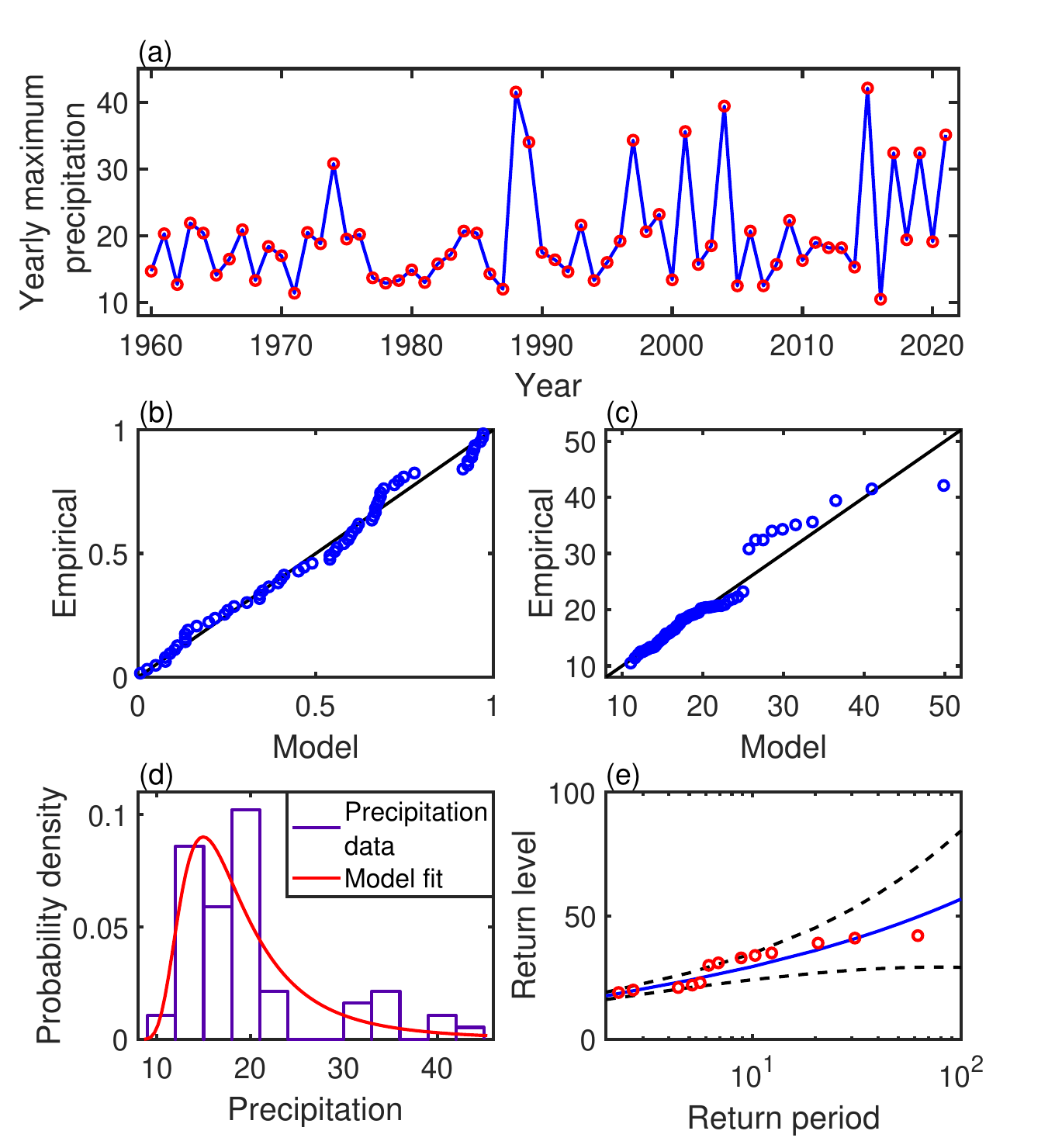}
 }
	\caption{(a) Time series of annual maximum precipitation events (red circles) during $1960$-$2021$.~(b) P-P plot along with straight line (back solid line). (c) Q-Q plot along with straight line (back solid line). (d) Histogram of extreme precipitation events fitted with probability density (red) of GEV distribution. (e) Return level from a statistical model (blue solid lines) associated with different return periods within $95\%$ confidence interval (denoted by back dashed line). }
	\label{fig3}
\end{figure*} 
\par Now, we discuss two terminologies, namely, the return level ($y_p$) and return period ($T$) that are usually used for statistical modeling, and essentially used here to derive an insight into the question of predictability of the occurrence of extreme precipitation.  Return level is a quantile that is defined as $P(y>y_p)=p$ and $y_p$ is a value that is expected to be exceeded on an average once every $\dfrac{1}{p}$ period. More precisely, it exceeds the extreme value in any particular period (year or season in our case) with probability $p$. Return period ($T$) is defined as $T=\dfrac{1}{p}$. Therefore, $P(y \le y_p)=G(y_p)=1-\dfrac{1}{T}$. Hence, the return level is calculated using an inverse distributional function (quantile function) defined as follows: \\	
\begin{equation}\label{eq.2}
y_p=G^{-1}\big(1-\dfrac{1}{T}\big)=
\begin{cases} 
\mu-\dfrac{\sigma}{\xi}\bigg[1-\{-log\big(1-\frac{1}{T}\big)\}^{-\xi}\bigg];\\~~~~~~~~~~~~~~~~~~~~~~~~~~~~~~~~ \xi \ne 0 \\
\mu-\sigma log\big[-log\big(1-\frac{1}{T}\big)\big];\\~~~~~~~~~~~~~~~~~~~~~~~~~~~~~~~~ \xi =0.
\end{cases}
\end{equation}

\section{\label{sec:Resut} Results: Precipitation extremes}
Firstly, we consider the whole period of years $[1960, 2021]$ taking each year as a block to characterize precipitation extremes. As a result, we obtain $62$ annual maxima of daily precipitation events, which we display in Fig.\ \ref{fig3}(a). The Augmented
Dickey-Fuller (ADF) test \cite{cheung1995lag} is used (see Appendix\ \ref{sec:appendix} for detail) for confirming the stationarity of the time series of annual maximum precipitation events (for our case, $p$-value = $0.03563<0.05$). The stationarity assumption must be satisfied to fit the distribution of these events with GEV distribution. We use probability plots and quantile plots to perform goodness-of-fit for model diagnostics of our dataset prior to fitting the distribution. A probability-probability (P-P) plot as well as quantile-quantile (Q-Q) plot compares the theoretical and empirical probability to see whether the fitting of a probability distribution represents a reasonable model. The P-P plot consists of the points
$\Big\{\Big({\hat G}(y_{i}),~\dfrac{i}{n+1}\Big):~i=1,2,\cdots,n \Big\}$ for ordered sample of independent observations $y_1\le y_2\le \dots \le y_n$ from a population with a distribution function $G$ and estimated distribution function ${\hat G}$. With an estimated distribution function ${\hat G}$, the Q-Q plot consists of the points
$\Big\{\Big({\hat G^{-1}}(y_{i}),~ y_i \Big):~i=1,2,\cdots,n \Big\}$. The P-P and Q-Q plots are displayed in Figs.\ \ref{fig3}(b) and \ref{fig3}(c), respectively, that show reasonable accuracy in fitting between the empirical and model distributions as points of these curves are scattered along the diagonal line (black). Now, we fit the histogram of annual maximum precipitations shown in Fig.\ \ref{fig3}(d) with probability density function (PDF) using GEV distribution as described by Eq.\ \ref{eq.1}. Thereby we estimate the parameters $\mu$, $\sigma$, and $\xi$ using the maximum likelihood estimation as presented in Table\ \ref{tab:multicol1}. When $\xi = 0.2894~(> 0)$, we can conclude that the probability distribution of maximum precipitation belongs to the Fr{\'e}chet family of distribution that denotes the heavy-tail of the distribution \cite{chattopadhyay2021modified}. 
The Kolmogorov-Smirnov (KS) test \cite{massey1951kolmogorov} (described in Appendix\ \ref{sec:appendix}) is another process of goodness-of-fit that is also performed on our dataset. A $p$-value $(=0.9769>0.05)$ indicates that the empirical distribution appropriately describes a GEV distribution.  
The return level plot against the return period of annual maximum precipitation is displayed in Fig.\ \ref{fig3}(e), which is used to predict the probability that the maximum precipitation exceeds a defined threshold limit. The estimation of return level corresponding to return periods of $10$-year and $100$-year are obtained using Eq.\ \ref{eq.2} as shown in Table\ \ref{tab:multicol1}. The black dashed lines in Fig.\ \ref{fig3}(e) represent the $95\%$ confidence interval of return levels. Red circles show the return level based on our observation.

{\centering
\begin{table*}[!hbt]
		
		\resizebox{1.0\textwidth}{!}{%
			\begin{tabular}{ |p{1.6cm}||p{1.6cm}|p{1.3cm}|p{1.3cm}||p{1.8cm}|p{1.6cm}||p{1.7cm}|p{1.5cm}|  }
				\hline
				\multicolumn{8}{|c|}{Parameter estimation of the GEV fitting with the standard errors, and return level estimation} \\
				\hline
				Station (station number) & Location $(\mu)$ (s.e.)& Scale $(\sigma)$ (s.e.)& Shape $(\xi)$ (s.e.) & negative log-likelihood & $p$-value for KS-test & 10-year return level & 100-year return level  \\
				\hline
				Bj{\o}rn{\o}ya  (SN99710)   & 15.9978 (0.6196) & 4.2406  (0.5203)  &  0.2894 (0.1183) & 197.7512 &  0.9769 &29.44779  & 56.818 \\
				\hline
			\end{tabular}%
		}
		\caption{Analysis of generalized extreme value distribution for annual maximum precipitation events of Bj{\o}rn{\o}ya, where s.e. denotes the standard error.}
		\label{tab:multicol1}
	\end{table*}
}

\begin{figure}
			\centerline{
	\includegraphics[width=0.7\textwidth]{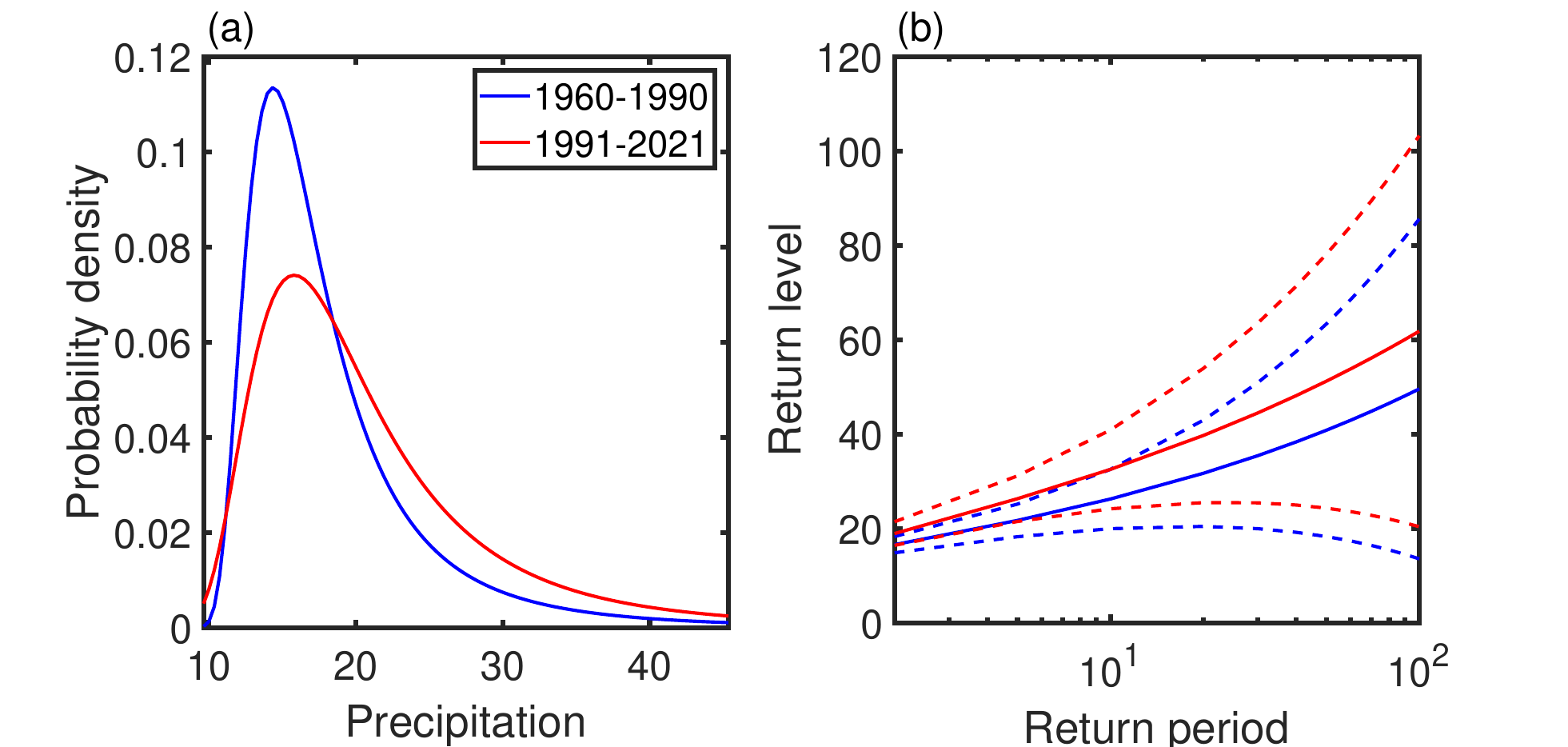}
			}
	\caption{(a) Estimated probability density function of generalized extreme value distribution. (b) The plot of return level (solid lines) for different return periods with $95\%$ confidence intervals (dashed lines). The blue line indicates the time period between $1960$ and $1990$ and for $[1991, 2021]$, the red color is used for depicting probability density as well as the return level plot. The estimated parameter values are $\mu=15.24, \sigma=3.385, \xi=0.308$ for the time period  $[1960, 1990]$ and $\mu=16.94, \sigma=5.118, \xi=0.256$ for the time period  $[1991, 2021]$.}
	\label{fig4}
\end{figure}

{\centering	
	\begin{table*}[!hbt]	
		\resizebox{1.0\textwidth}{!}{%
			
			\begin{tabular}{ |p{1.4cm}||p{1.7cm}|p{1.6cm}|p{1.6cm}||p{1.8cm}|p{1.4cm}||p{1.4cm}|p{1.4cm}| }
				\hline
				\multicolumn{8}{|c|}{Parameter estimation of the GEV fitting with the standard errors and return level estimation} \\
				\hline
				Season & Location $(\mu)$ & Scale $(\sigma)$ & Shape $(\xi)$ & negative log-likelihood & $p$-value for KS-test & $10$-year return level & $100$-year return level  \\
				\hline
				Spring & 7.9505  (0.4995)& 3.2829  (0.3847)  & 0.0527  (0.1343) &173.7239 &  0.9973  & 15.796717 & 25.036279  \\	        
				\hline
				Summer & 11.2199 (0.703841)& 4.97904 (0.51975) &  0.0768059  (0.0875837)&200.086 &  0.9987 & 23.45120 & 38.69230  \\
				\hline
				Autumn &11.8318 (0.557175) & 3.97918 (0.406001) & 0.0703106  (0.0795766)&185.797 &  0.9983 & 21.53269 & 33.44324  \\	        
				\hline
				Winter & 9.54765 (0.549231) & 3.88695 (0.43483) & 0.215446 (0.0918661) &189.695  & 0.9977 & 20.80330 & 40.11102  \\
				\hline
			\end{tabular}%
		}
		\caption{Analysis of generalized extreme value distribution for seasonal maximum precipitation events of Bj{\o}rn{\o}ya}.
		\label{tab:multicol2}
	\end{table*}
}

\par In general, a climate time scale is a period of $30$ years \cite{ghosh2012lack}. Therefore, we divide the 62 years, $(1960$-$2021)$ into two equal size halves $(1960$-$1990)$ and $(1991$-$2021)$. We make a comparative analysis of the frequency of extreme precipitations for the two consecutive climate timescales, separately, and plot the probability density functions in Fig.\ \ref{fig4}(a) for the two time periods using the model of GEV distribution. The approximate values of the shape parameter $\xi$ are $0.308 \; (>0)$ and $0.256 \; (>0)$ for the respective halves $[1960, 1990]$ and $[1991, 2021]$. This signifies that both the distributions still belong to the Fr{\'e}chet family of distribution and hence we conclude that both the distributions are the heavy-tailed distribution.
From a visual inspection of the annual precipitation maxima plot in Fig.\ \ref{fig3}(a), it is noticed that the frequency of precipitation extremes with high magnitude has increased during the second half $[1991, 2021]$ of $31$ years compared to the first half $[1960, 1990]$ time period. This is reflected in the higher thickness of the tail of the probability density for the second half compared to the first one as shown in Fig.\ \ref{fig4}(a).
The return levels (solid red and blue lines) of the two climate timescales are plotted against the return period with bounds of $95\%$ confidence interval (dashed lines) in Fig.\ \ref{fig4}(b). They illustrate the increasing trend in the occurrence of precipitation extremes in the second half of $31$ years (solid red lines) than the occurrences in the first half, $(1960$-$1990)$ (solid blue lines). This figure concludes the increasing trend in precipitation extremes in the future and these events are increasing at a faster rate for the last $31$ years compared to the $(1960$-$1990)$ time period.

\begin{figure*}
	\centerline{
		\includegraphics[width=1.15\textwidth]{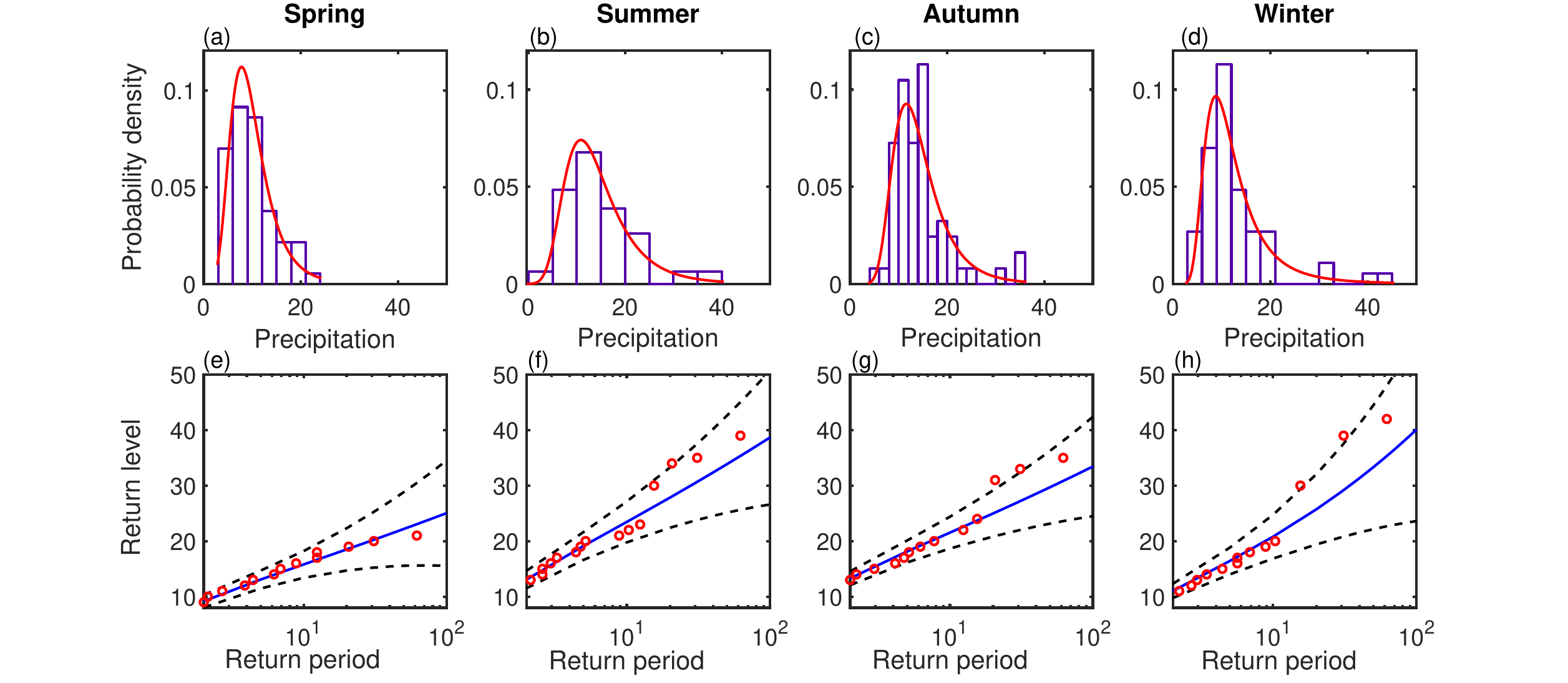}
	}
	\caption{(a)-(d) Histograms (blue) of seasonal data of precipitation extremes fitted with probability densities (red) of GEV distributions for four seasons, i.e., spring, summer, autumn, and winter respectively.~(e)-(h) Return level plots (blue solid lines) from statistical models against  return periods within $95\%$ confidence interval (black dashed lines). Return levels are depicted from empirical data (red circles).}
	\label{fig5}
\end{figure*}

\begin{figure*}
	\centerline{
		\includegraphics[width=1.15\textwidth]{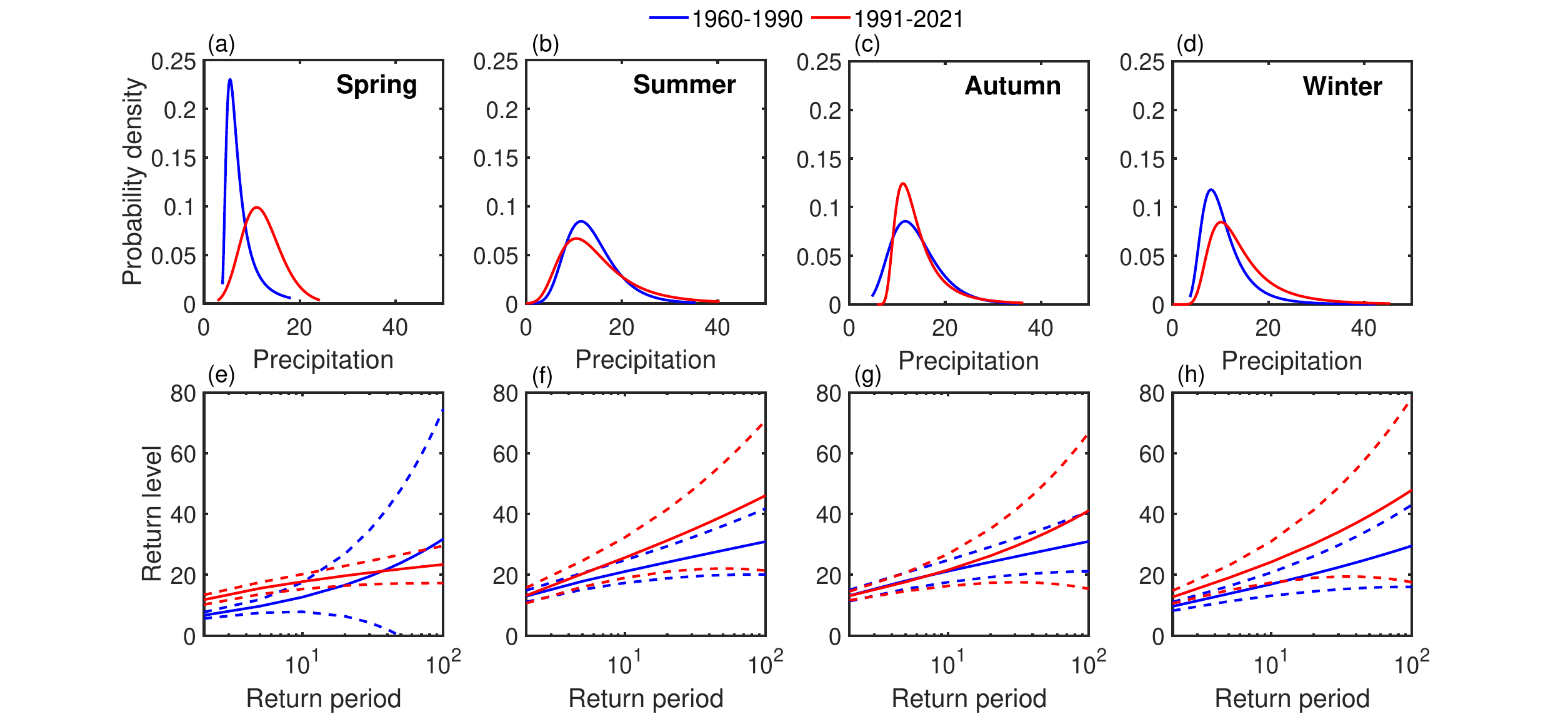}
	}
	\caption{(a)-(d) Probability density functions of generalized extreme value distributions are shown for four seasons spring, summer autumn, and winter. (e)-(h) The return levels (solid lines) are  delineated for different return periods with $95\%$ confidence intervals (dashed lines) corresponding to probability densities. The blue and red lines are used for the time period $[1960, 1990]$ and $[1991, 2021]$, respectively.}
	\label{fig6}
\end{figure*}

\par The seasonal variation is now looked into when we consider four seasons, separately, each season as being a block period, (i) spring (March, April, May), (ii) summer (June, July, August), (iii) autumn (September, October, November) and (iv) winter (December, January, February). Four datasets covering each season of the years  $1960$-$2021$ contain $62$ seasonal maximum precipitation events. In order to derive any statistical model of the precipitation extremes, we apply EVT once again, but for each of the four seasons we take it as a separate entity and create four seasonal data sets. Figures\ \ref{fig5}(a)-(d) show the empirical probability density plots for four seasons fitted with PDF (red line) of GEV distribution model. The return levels (blue line) as derived using Eq.\ \ref{eq.2} are plotted against the return periods with $95\%$ confidence levels (black dashed lines)  as shown in Figs.\ \ref{fig5}(e)-(h) for the four seasons. We use the maximum likelihood estimation method to estimate the parameters $\mu$, $\sigma$ and $\xi$. We perform goodness-of-fit using the KS test for the four seasons and check that the statistical model is well-fitted with the empirical probability density plot. We accumulate the estimated values of $\mu$, $\sigma$, and $\xi$, log-likelihood, p-values for KS test, $10$-year return level, $100$-year return level in Table.\ \ref{tab:multicol2} for four seasons. We conclude that all the probability distribution functions from the statistical model fall into the category of Fr\'echet distribution, i.e., heavy-tailed distribution, and these probability distribution functions are aptly fitted with available data sets for all four seasons.

\par Now we do the same analysis on the four seasons for two time periods of 31 years' block, separately.  Our target is to determine the seasonal variability of precipitation extremes over the two successive climate periods. The probability density functions of GEV distribution are drawn in Figs.\ \ref{fig6}(a)-(d) for the four seasons (spring, summer, autumn, and winter, respectively) of the two climate timescales. The return levels (solid lines) against return periods with $95\%$ confidence intervals (dashed lines) are plotted in Figs.\ \ref{fig6}(e)-(h). Firstly, we take a closer look at Fig.\ \ref{fig6}(a) which presents PDF of GEV distribution of the two separate data sets consisting of maximum precipitation events in spring for the years $[1960, 1990]$ (blue line) and $[1991, 2021]$ (red line). The estimated values of the shape parameter of GEV distributions are $0.438$ and $-0.133$, respectively for the season spring. It appears from the shape parameters that the probability distribution belongs to the Fr{\'e}chet family  (blue line) for the first time period while for the second one, it is a member of the Weibull family of distribution (red line). The tails of probability density plots confirm that the frequency of high amplitude precipitation extremes is higher for the second time period than it is for the first one. But, nature of the family of probability distribution has changed. We also notice from the corresponding return level against the return period plot in Fig.\ \ref{fig6}(e) that for a certain range of return period, the curve of the return level for the second period (red line) dominates over the spring of the first period (blue line). However, it becomes submissive after a certain value of the return period, and the curve of the return level of the first period dominates. For the time periods $[1960, 1990]$ and $[1991, 2021]$, the estimated values of shape parameter $\xi$ of GEV distributions are, respectively, $-0.01095$ and $0.13$ for summer precipitation extremes  and $-0.00916$ and $0.285$ for fall precipitation extremes. For both summer and autumn, the probability distributions belong to the Weibull family for the first time period and the Fr{\'e}chet family rather than say, heavy-tailed probability distribution for the second period. The tail of the probability density plot (blue line) for $[1960, 1990]$ is slightly heavier than the same plot (red line) for $[1991, 2021]$ in the summer precipitation extremes, as seen in Fig.\ \ref{fig6}(b). The return level (blue line) for the first time period completely outweighs the return level  (red line) for the second time period, as shown in Fig.\ \ref{fig6}(f). On the other hand, for the autumn (or, fall), the tail of the probability density plot (blue line) for $[1960, 1990]$ is almost similar to the density plot (red) for $[1991, 2021]$, which we observe from Fig.\ \ref{fig6}(c). The return level during the first half follows the second half for a wide range of return periods as shown in Fig.\ \ref{fig6}(g). For higher return periods, the return level during the $[1960, 1990]$ becomes submissive by the return level of $[1991, 2021]$. We depict the probability density for the winter extreme precipitation events in Fig.\ \ref{fig6}(d). For the winter periods in $1960$-$1990$ and $1991$-$2021$, the estimated values of the shape parameter ($\xi$) are $0.154 \; (>0)$ and $0.235 \; (>0)$, respectively. Consequently, in the winter season for both periods, the probability distributions are members of the Fr{\'e}chet family. In comparison, the frequency of extreme precipitation events of high magnitude has risen in the winter of $[1991, 2021]$ from the winter in $[1960, 1990]$ as seen in Fig.\ \ref{fig6}(d). The return level plot against the return period in Fig.\ \ref{fig6}(h) also confirms the faster rate of increase in the frequency of extreme precipitation events, which has been occurring more frequently in the last $31$ years than in the years between $1960$ and $1990$. Overall, the amount of heavy precipitation has risen over the four seasons of the past $31$ years compared to the seasons between $1960$ and $1990$, especially during the winter. Also, for each seasonal analysis of the return level plot, we can conclude that the frequency of precipitation extremes will be increasing in the future.

\begin{figure}
	\centerline{
\includegraphics[width=0.48\textwidth]{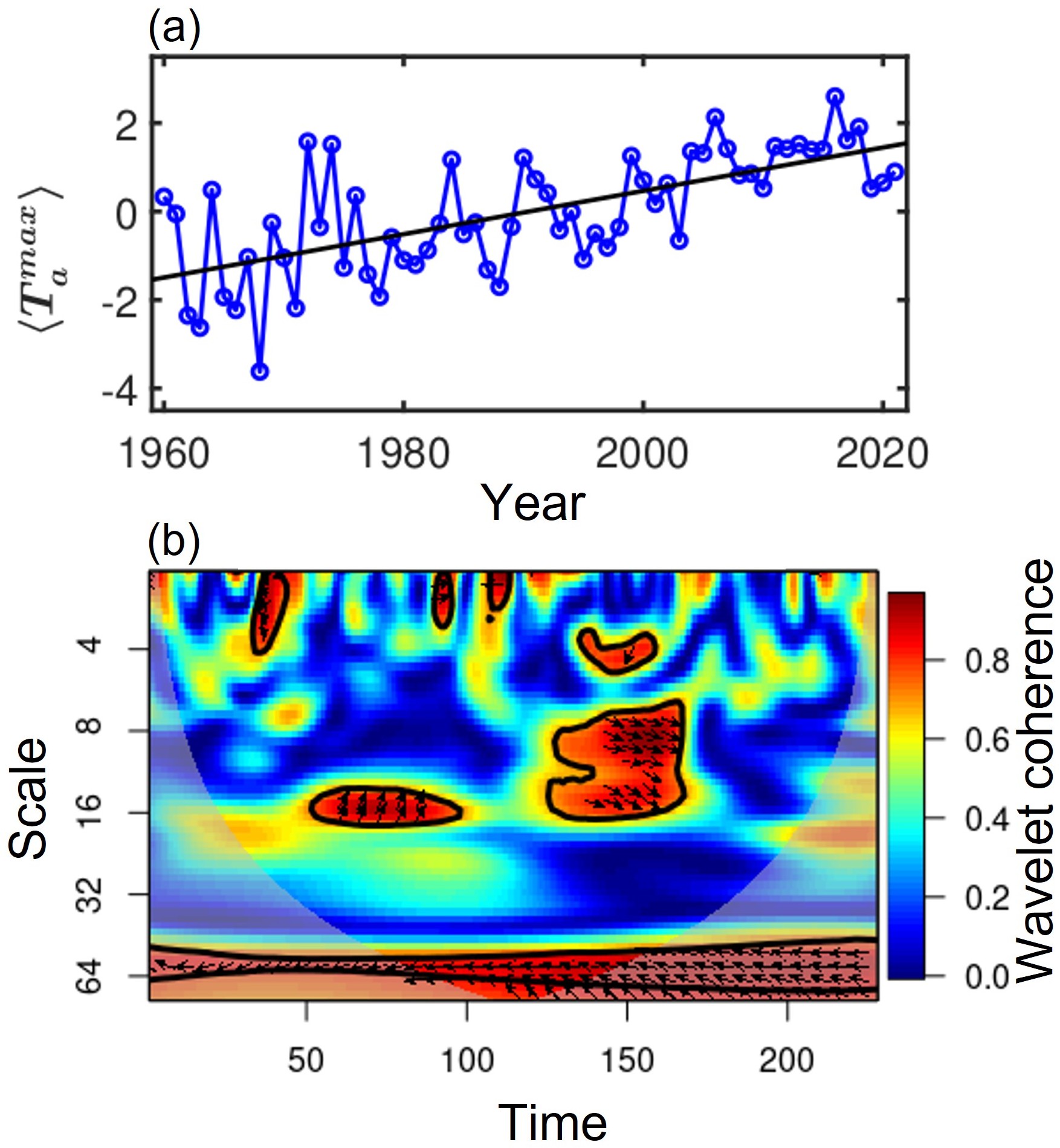}
	}
\caption{(a) Time series (blue) of the yearly mean of temperature anomaly $\langle T^{max}_a \rangle$. It exhibits a linear increasing trend (back line) with an estimated slope $0.04899^o$C/year for the whole period $[1960, 2021]$. (b) Wavelet coherence plot between precipitation extremes and yearly temperature anomaly ($T^{max}_a$) corresponding to the day of occurrence of precipitation extremes. The horizontal axis shows the time of observation, and the vertical axis shows the scale. The color bar indicates the strength of coherence, weaker  (blue) to stronger (red) colors indicate an increasingly significant interrelationship between extreme precipitation events and temperature. The arrows represent the direction of the relationship. Time and Scale domain representations are obtained by wavelet decomposition. The area beyond the white-shaded region indicates a lack of association between the variables.}
	\label{fig7}
\end{figure}

\par Finally, we look into the temperature data for the extended period of $1960-2021$ and search if there exists any dependency between yearly maximum precipitation events and the yearly mean of daily maximum temperature anomaly. The annual mean of daily maximum temperature anomaly gradually rises over the years as shown in Fig.\ \ref{fig7}(a). This is linearly fitted with a slope $0.04899$. In order to calculate the temperature anomaly, we use the average daily temperature for the period between $1960$ and $2021$ as the base value. The temperature anomaly is then determined by deducting the base value from the daily temperature. After getting the temperature anomaly ($T^{max}_a$), we compute the average over a year.
Since there is an increasing trend for annual average maximum shown in Fig.\ \ref{fig7}(a) and annual average precipitation highlighted in Fig.\ \ref{fig7}(a), we try to find if there is any dependency between precipitation extremes and the temperature anomaly. 
We use wavelet transform coherence \cite {rehana2022precipitation, panja2023ensemble} to examine how the annual mean of daily maximum temperature anomaly and precipitation extremes are related. From Fig.\ \ref{fig7}(b), we realize that precipitation extremes are sometimes related to the yearly average of temperature anomaly. The color bar indicates the strength of coherence:  a strong association (red color) between the two time series of extreme precipitation and the maximum temperature anomaly of that day does exist in a reasonably significant region of the time-scale plane. 
The arrows on the plot show the direction of the relationship between the variables such that the right-down (right-up) or left-up(left-down) direction reveals that the precipitation extremes (temperature anomaly) are the major cause of the temperature anomaly (precipitation extremes).

\section{\label{sec:discussion}Discussions and Conclusion}
We have collected daily precipitation data in Bear Island over a period of $62$ years spanning from $1960$ to $2021$ and then attempted statistical analysis of the precipitation extremes and thereby try to understand occurrences of precipitation extremes and their variability by variation of climate timescales. Also, we have tried to find if the dataset of yearly precipitation extremes has any connection to the yearly mean of the daily maximum temperature of the island. We used the block maxima method to identify and characterize precipitation extremes, considering both the annual and seasonal blocks, separately. We fitted the recorded precipitation extreme data with the Generalized extreme value distribution model and estimated the associated parameter values to classify them as belonging to either the Fr\'echet or the Weibull family of distributions. We have generated return level plots for different return periods, providing valuable insights on the predictability of extreme precipitation extremes in the future. Our analysis has revealed climate variability of precipitation extremes in Bear Island over the climate timescales. By dividing the study period into two distinct intervals of years, $[1960, 1990]$ and $[1991, 2021]$, we have observed an increase in the frequency of high-amplitude precipitation extremes during the latter 31 years. While our findings suggest that Arctic warming has influenced extreme precipitation in Bear Island (Bjørnøya), further research is required to establish a consensus, particularly, in other Arctic regions. Nonetheless, this study serves as an early warning system for climate conservationists and provides a valuable foundation for future investigations on climate change and its impacts. 
\par Our preliminary research based on real data contributes to the understanding of extreme precipitation patterns in Bear Island and highlights the importance of considering extreme events in climate conservation efforts, in general. This research effort aimed at joining hands with the policymakers and stakeholders if our insights based on empirical data anyway help implementing appropriate strategies for climate change adaptation and mitigation in this location. However, ongoing research is crucial to deepen our understanding of the complex dynamics between Arctic warming, extreme events, and their potential consequences. By expanding the scope of our study to other Arctic regions, we can further refine our understanding of climate change impacts. In conclusion, this study at present sheds light on the extreme precipitation patterns in Bear Island and demonstrates the utility of statistical modeling and extreme value theory in characterizing and analyzing precipitation extremes.

\bibliography{precipitation_bib}

\begin{thebibliography}{10}
\urlstyle{rm}
\expandafter\ifx\csname url\endcsname\relax
  \def\url#1{\texttt{#1}}\fi
\expandafter\ifx\csname urlprefix\endcsname\relax\def\urlprefix{URL }\fi
\expandafter\ifx\csname doiprefix\endcsname\relax\def\doiprefix{DOI: }\fi
\providecommand{\bibinfo}[2]{#2}
\providecommand{\eprint}[2][]{\url{#2}}

\bibitem{albeverio2006extreme}
\bibinfo{author}{Albeverio, S.}, \bibinfo{author}{Jentsch, V.} \& \bibinfo{author}{Kantz, H.}
\newblock \emph{\bibinfo{title}{Extreme events in nature and society}} (\bibinfo{publisher}{Springer}, \bibinfo{address}{London}, \bibinfo{year}{2006}).

\bibitem{portner2022climate}
\bibinfo{author}{P{\"o}rtner, H.-O.} \emph{et~al.}
\newblock \emph{\bibinfo{title}{Climate change 2022: Impacts, adaptation and vulnerability}} (\bibinfo{publisher}{IPCC Geneva}, \bibinfo{address}{Switzerland}, \bibinfo{year}{2022}).

\bibitem{kangalawe2017climate}
\bibinfo{author}{Kangalawe, R.~Y.}, \bibinfo{author}{Mung’ong’o, C.~G.}, \bibinfo{author}{Mwakaje, A.~G.}, \bibinfo{author}{Kalumanga, E.} \& \bibinfo{author}{Yanda, P.~Z.}
\newblock \bibinfo{journal}{\bibinfo{title}{Climate change and variability impacts on agricultural production and livelihood systems in western tanzania}}.
\newblock {\emph{\JournalTitle{Climate and Development}}} \textbf{\bibinfo{volume}{9}}, \bibinfo{pages}{202--216} (\bibinfo{year}{2017}).

\bibitem{huq2004mainstreaming}
\bibinfo{author}{Huq, S.} \emph{et~al.}
\newblock \bibinfo{journal}{\bibinfo{title}{Mainstreaming adaptation to climate change in least developed countries (ldcs)}}.
\newblock {\emph{\JournalTitle{Climate Policy}}} \textbf{\bibinfo{volume}{4}}, \bibinfo{pages}{25--43} (\bibinfo{year}{2004}).

\bibitem{vihma2016atmospheric}
\bibinfo{author}{Vihma, T.} \emph{et~al.}
\newblock \bibinfo{journal}{\bibinfo{title}{The atmospheric role in the arctic water cycle: A review on processes, past and future changes, and their impacts}}.
\newblock {\emph{\JournalTitle{Journal of Geophysical Research: Biogeosciences}}} \textbf{\bibinfo{volume}{121}}, \bibinfo{pages}{586--620} (\bibinfo{year}{2016}).

\bibitem{walsh2020extreme}
\bibinfo{author}{Walsh, J.~E.} \emph{et~al.}
\newblock \bibinfo{journal}{\bibinfo{title}{Extreme weather and climate events in northern areas: A review}}.
\newblock {\emph{\JournalTitle{Earth-Science Reviews}}} \textbf{\bibinfo{volume}{209}}, \bibinfo{pages}{103324} (\bibinfo{year}{2020}).

\bibitem{bintanja2020strong}
\bibinfo{author}{Bintanja, R.} \emph{et~al.}
\newblock \bibinfo{journal}{\bibinfo{title}{Strong future increases in arctic precipitation variability linked to poleward moisture transport}}.
\newblock {\emph{\JournalTitle{Science advances}}} \textbf{\bibinfo{volume}{6}}, \bibinfo{pages}{eaax6869} (\bibinfo{year}{2020}).

\bibitem{przybylak2003climate}
\bibinfo{author}{Przybylak, R.}, \bibinfo{author}{Sadourny, R.} \& \bibinfo{author}{Mysak, L.~A.}
\newblock \emph{\bibinfo{title}{The climate of the Arctic}} (\bibinfo{publisher}{Springer}, \bibinfo{year}{2003}).

\bibitem{box2019key}
\bibinfo{author}{Box, J.~E.} \emph{et~al.}
\newblock \bibinfo{journal}{\bibinfo{title}{Key indicators of arctic climate change: 1971--2017}}.
\newblock {\emph{\JournalTitle{Environmental Research Letters}}} \textbf{\bibinfo{volume}{14}}, \bibinfo{pages}{045010} (\bibinfo{year}{2019}).

\bibitem{landrum2020extremes}
\bibinfo{author}{Landrum, L.} \& \bibinfo{author}{Holland, M.~M.}
\newblock \bibinfo{journal}{\bibinfo{title}{Extremes become routine in an emerging new arctic}}.
\newblock {\emph{\JournalTitle{Nature Climate Change}}} \textbf{\bibinfo{volume}{10}}, \bibinfo{pages}{1108--1115} (\bibinfo{year}{2020}).

\bibitem{owczarek2020post}
\bibinfo{author}{Owczarek, P.}, \bibinfo{author}{Opa{\l}a-Owczarek, M.} \& \bibinfo{author}{Miga{\l}a, K.}
\newblock \bibinfo{journal}{\bibinfo{title}{Post-1980s shift in the sensitivity of tundra vegetation to climate revealed by the first dendrochronological record from bear island (bj{\o}rn{\o}ya), western barents sea}}.
\newblock {\emph{\JournalTitle{Environmental Research Letters}}} \textbf{\bibinfo{volume}{16}}, \bibinfo{pages}{014031} (\bibinfo{year}{2020}).

\bibitem{ghosh2012lack}
\bibinfo{author}{Ghosh, S.}, \bibinfo{author}{Das, D.}, \bibinfo{author}{Kao, S.-C.} \& \bibinfo{author}{Ganguly, A.~R.}
\newblock \bibinfo{journal}{\bibinfo{title}{Lack of uniform trends but increasing spatial variability in observed indian rainfall extremes}}.
\newblock {\emph{\JournalTitle{Nature Climate Change}}} \textbf{\bibinfo{volume}{2}}, \bibinfo{pages}{86--91} (\bibinfo{year}{2012}).

\bibitem{santos2016estimating}
\bibinfo{author}{Santos, E.~B.}, \bibinfo{author}{Lucio, P.~S.} \& \bibinfo{author}{Santos~e Silva, C.~M.}
\newblock \bibinfo{journal}{\bibinfo{title}{Estimating return periods for daily precipitation extreme events over the brazilian amazon}}.
\newblock {\emph{\JournalTitle{Theoretical and Applied Climatology}}} \textbf{\bibinfo{volume}{126}}, \bibinfo{pages}{585--595} (\bibinfo{year}{2016}).

\bibitem{ngailo2016modelling}
\bibinfo{author}{Ngailo, T.~J.}, \bibinfo{author}{Reuder, J.}, \bibinfo{author}{Rutalebwa, E.}, \bibinfo{author}{Nyimvua, S.} \& \bibinfo{author}{Mesquita, M.}
\newblock \bibinfo{journal}{\bibinfo{title}{Modelling of extreme maximum rainfall using extreme value theory for tanzania}}.
\newblock {\emph{\JournalTitle{International Journal of Scientific and Innovative Mathematical Research}}} \textbf{\bibinfo{volume}{4}}, \bibinfo{pages}{34--45} (\bibinfo{year}{2016}).

\bibitem{yozgatligil2018extreme}
\bibinfo{author}{Yozgatl{\i}gil, C.} \& \bibinfo{author}{T{\"u}rke{\c{s}}, M.}
\newblock \bibinfo{journal}{\bibinfo{title}{Extreme value analysis and forecasting of maximum precipitation amounts in the western black sea subregion of turkey}}.
\newblock {\emph{\JournalTitle{International Journal of Climatology}}} \textbf{\bibinfo{volume}{38}}, \bibinfo{pages}{5447--5458} (\bibinfo{year}{2018}).

\bibitem{tabari2021extreme}
\bibinfo{author}{Tabari, H.}
\newblock \bibinfo{journal}{\bibinfo{title}{Extreme value analysis dilemma for climate change impact assessment on global flood and extreme precipitation}}.
\newblock {\emph{\JournalTitle{Journal of Hydrology}}} \textbf{\bibinfo{volume}{593}}, \bibinfo{pages}{125932} (\bibinfo{year}{2021}).

\bibitem{murukeshcharacteristics}
\bibinfo{author}{Athulya, R.}, \bibinfo{author}{Nuncio, M.}, \bibinfo{author}{Chatterjee, S.} \& \bibinfo{author}{Vidya, P.}
\newblock \bibinfo{journal}{\bibinfo{title}{Characteristics of the mean and extreme precipitation in ny Ålesund, arctic}}.
\newblock {\emph{\JournalTitle{Atmospheric Research}}} \textbf{\bibinfo{volume}{295}}, \bibinfo{pages}{106989} (\bibinfo{year}{2023}).

\bibitem{nuncio2023hails}
\bibinfo{author}{Nuncio, M.} \emph{et~al.}
\newblock \bibinfo{journal}{\bibinfo{title}{Hails in ny Ålesund, svalbard-atmospheric vertical structure and dependence on circulation}}.
\newblock {\emph{\JournalTitle{Natural Hazards}}} \textbf{\bibinfo{volume}{117}}, \bibinfo{pages}{1365--1380} (\bibinfo{year}{2023}).

\bibitem{gnedenko1943distribution}
\bibinfo{author}{Gnedenko, B.}
\newblock \bibinfo{journal}{\bibinfo{title}{Sur la distribution limite du terme maximum d'une serie aleatoire}}.
\newblock {\emph{\JournalTitle{Annals of Mathematics}}} \bibinfo{pages}{423--453} (\bibinfo{year}{1943}).

\bibitem{kotz2000extreme}
\bibinfo{author}{Kotz, S.} \& \bibinfo{author}{Nadarajah, S.}
\newblock \emph{\bibinfo{title}{Extreme value distributions: theory and applications}} (\bibinfo{publisher}{Imperial College Press}, \bibinfo{address}{London}, \bibinfo{year}{2000}).

\bibitem{bali2003generalized}
\bibinfo{author}{Bali, T.~G.}
\newblock \bibinfo{journal}{\bibinfo{title}{The generalized extreme value distribution}}.
\newblock {\emph{\JournalTitle{Economics Letters}}} \textbf{\bibinfo{volume}{79}}, \bibinfo{pages}{423--427} (\bibinfo{year}{2003}).

\bibitem{ghil2011extreme}
\bibinfo{author}{Ghil, M.} \emph{et~al.}
\newblock \bibinfo{journal}{\bibinfo{title}{Extreme events: dynamics, statistics and prediction}}.
\newblock {\emph{\JournalTitle{Nonlinear Processes in Geophysics}}} \textbf{\bibinfo{volume}{18}}, \bibinfo{pages}{295--350} (\bibinfo{year}{2011}).

\bibitem{engeland2004practical}
\bibinfo{author}{Engeland, K.}, \bibinfo{author}{Hisdal, H.} \& \bibinfo{author}{Frigessi, A.}
\newblock \bibinfo{journal}{\bibinfo{title}{Practical extreme value modelling of hydrological floods and droughts: a case study}}.
\newblock {\emph{\JournalTitle{Extremes}}} \textbf{\bibinfo{volume}{7}}, \bibinfo{pages}{5--30} (\bibinfo{year}{2004}).

\bibitem{poon2004extreme}
\bibinfo{author}{Poon, S.-H.}, \bibinfo{author}{Rockinger, M.} \& \bibinfo{author}{Tawn, J.}
\newblock \bibinfo{journal}{\bibinfo{title}{Extreme value dependence in financial markets: Diagnostics, models, and financial implications}}.
\newblock {\emph{\JournalTitle{The Review of Financial Studies}}} \textbf{\bibinfo{volume}{17}}, \bibinfo{pages}{581--610} (\bibinfo{year}{2004}).

\bibitem{mcneil1997estimating}
\bibinfo{author}{McNeil, A.~J.}
\newblock \bibinfo{journal}{\bibinfo{title}{Estimating the tails of loss severity distributions using extreme value theory}}.
\newblock {\emph{\JournalTitle{ASTIN Bulletin: The Journal of the IAA}}} \textbf{\bibinfo{volume}{27}}, \bibinfo{pages}{117--137} (\bibinfo{year}{1997}).

\bibitem{coles2001}
\bibinfo{author}{Coles, S.}
\newblock \emph{\bibinfo{title}{An Introduction to Statistical Modeling of Extreme Values}} (\bibinfo{publisher}{Springer-Verlag}, \bibinfo{address}{London}, \bibinfo{year}{2001}), \bibinfo{edition}{1} edn.

\bibitem{lucarini2016extremes}
\bibinfo{author}{Lucarini, V.} \emph{et~al.}
\newblock \emph{\bibinfo{title}{Extremes and recurrence in dynamical systems}} (\bibinfo{publisher}{John Wiley \& Sons}, \bibinfo{address}{New Jersey}, \bibinfo{year}{2016}).

\bibitem{torrence1998practical}
\bibinfo{author}{Torrence, C.} \& \bibinfo{author}{Compo, G.~P.}
\newblock \bibinfo{journal}{\bibinfo{title}{A practical guide to wavelet analysis}}.
\newblock {\emph{\JournalTitle{Bulletin of the American Meteorological society}}} \textbf{\bibinfo{volume}{79}}, \bibinfo{pages}{61--78} (\bibinfo{year}{1998}).

\bibitem{worslex2001geological}
\bibinfo{author}{Worslex, D.} \emph{et~al.}
\newblock \bibinfo{journal}{\bibinfo{title}{The geological evolution of bi{\o}rn{\o}ya, arctic norway: implications for the barents shelf.}}
\newblock {\emph{\JournalTitle{Norwegian Journal of Geology/Norsk Geologisk Forening}}} \textbf{\bibinfo{volume}{81}} (\bibinfo{year}{2001}).

\bibitem{schar2016percentile}
\bibinfo{author}{Sch{\"a}r, C.} \emph{et~al.}
\newblock \bibinfo{journal}{\bibinfo{title}{Percentile indices for assessing changes in heavy precipitation events}}.
\newblock {\emph{\JournalTitle{Climatic Change}}} \textbf{\bibinfo{volume}{137}}, \bibinfo{pages}{201--216} (\bibinfo{year}{2016}).

\bibitem{mcphillips2018defining}
\bibinfo{author}{McPhillips, L.~E.} \emph{et~al.}
\newblock \bibinfo{journal}{\bibinfo{title}{Defining extreme events: A cross-disciplinary review}}.
\newblock {\emph{\JournalTitle{Earth's Future}}} \textbf{\bibinfo{volume}{6}}, \bibinfo{pages}{441--455} (\bibinfo{year}{2018}).

\bibitem{mishra2020routes}
\bibinfo{author}{Mishra, A.} \emph{et~al.}
\newblock \bibinfo{journal}{\bibinfo{title}{Routes to extreme events in dynamical systems: Dynamical and statistical characteristics}}.
\newblock {\emph{\JournalTitle{Chaos: An Interdisciplinary Journal of Nonlinear Science}}} \textbf{\bibinfo{volume}{30}}, \bibinfo{pages}{063114} (\bibinfo{year}{2020}).

\bibitem{gilbert1987statistical}
\bibinfo{author}{Gilbert, R.~O.}
\newblock \emph{\bibinfo{title}{Statistical methods for environmental pollution monitoring}} (\bibinfo{publisher}{Van Nostrand Reinhold Company}, \bibinfo{address}{New York}, \bibinfo{year}{1987}).

\bibitem{bunde2005long}
\bibinfo{author}{Bunde, A.}, \bibinfo{author}{Eichner, J.~F.}, \bibinfo{author}{Kantelhardt, J.~W.} \& \bibinfo{author}{Havlin, S.}
\newblock \bibinfo{journal}{\bibinfo{title}{Long-term memory: A natural mechanism for the clustering of extreme events and anomalous residual times in climate records}}.
\newblock {\emph{\JournalTitle{Physical Review Letters}}} \textbf{\bibinfo{volume}{94}}, \bibinfo{pages}{048701} (\bibinfo{year}{2005}).

\bibitem{heeger2000poisson}
\bibinfo{author}{Heeger, D.} \emph{et~al.}
\newblock \bibinfo{journal}{\bibinfo{title}{Poisson model of spike generation}}.
\newblock {\emph{\JournalTitle{Handout, University of Standford}}} \textbf{\bibinfo{volume}{5}}, \bibinfo{pages}{76} (\bibinfo{year}{2000}).

\bibitem{chakraborty2022searching}
\bibinfo{author}{Chakraborty, T.}, \bibinfo{author}{Chattopadhyay, S.}, \bibinfo{author}{Das, S.}, \bibinfo{author}{Kumar, U.} \& \bibinfo{author}{Senthilnath, J.}
\newblock \bibinfo{journal}{\bibinfo{title}{Searching for heavy-tailed probability distributions for modeling real-world complex networks}}.
\newblock {\emph{\JournalTitle{IEEE Access}}} \textbf{\bibinfo{volume}{10}}, \bibinfo{pages}{115092--115107} (\bibinfo{year}{2022}).

\bibitem{cheung1995lag}
\bibinfo{author}{Cheung, Y.-W.} \& \bibinfo{author}{Lai, K.~S.}
\newblock \bibinfo{journal}{\bibinfo{title}{Lag order and critical values of the augmented dickey--fuller test}}.
\newblock {\emph{\JournalTitle{Journal of Business \& Economic Statistics}}} \textbf{\bibinfo{volume}{13}}, \bibinfo{pages}{277--280} (\bibinfo{year}{1995}).

\bibitem{chattopadhyay2021modified}
\bibinfo{author}{Chattopadhyay, S.}, \bibinfo{author}{Chakraborty, T.}, \bibinfo{author}{Ghosh, K.} \& \bibinfo{author}{Das, A.~K.}
\newblock \bibinfo{journal}{\bibinfo{title}{Modified lomax model: A heavy-tailed distribution for fitting large-scale real-world complex networks}}.
\newblock {\emph{\JournalTitle{Social Network Analysis and Mining}}} \textbf{\bibinfo{volume}{11}}, \bibinfo{pages}{43} (\bibinfo{year}{2021}).

\bibitem{massey1951kolmogorov}
\bibinfo{author}{Massey~Jr, F.~J.}
\newblock \bibinfo{journal}{\bibinfo{title}{The kolmogorov-smirnov test for goodness of fit}}.
\newblock {\emph{\JournalTitle{Journal of the American Statistical Association}}} \textbf{\bibinfo{volume}{46}}, \bibinfo{pages}{68--78} (\bibinfo{year}{1951}).

\bibitem{rehana2022precipitation}
\bibinfo{author}{Rehana, S.}, \bibinfo{author}{Yeleswarapu, P.}, \bibinfo{author}{Basha, G.} \& \bibinfo{author}{Munoz-Arriola, F.}
\newblock \bibinfo{journal}{\bibinfo{title}{Precipitation and temperature extremes and association with large-scale climate indices: An observational evidence over india}}.
\newblock {\emph{\JournalTitle{Journal of Earth System Science}}} \textbf{\bibinfo{volume}{131}}, \bibinfo{pages}{170} (\bibinfo{year}{2022}).

\bibitem{panja2023ensemble}
\bibinfo{author}{Panja, M.} \emph{et~al.}
\newblock \bibinfo{journal}{\bibinfo{title}{An ensemble neural network approach to forecast dengue outbreak based on climatic condition}}.
\newblock {\emph{\JournalTitle{Chaos, Solitons \& Fractals}}} \textbf{\bibinfo{volume}{167}}, \bibinfo{pages}{113124} (\bibinfo{year}{2023}).

\bibitem{gocic2013analysis}
\bibinfo{author}{Gocic, M.} \& \bibinfo{author}{Trajkovic, S.}
\newblock \bibinfo{journal}{\bibinfo{title}{Analysis of changes in meteorological variables using mann-kendall and sen's slope estimator statistical tests in serbia}}.
\newblock {\emph{\JournalTitle{Global and Planetary Change}}} \textbf{\bibinfo{volume}{100}}, \bibinfo{pages}{172--182} (\bibinfo{year}{2013}).

\bibitem{fuller2009introduction}
\bibinfo{author}{Fuller, W.~A.}
\newblock \emph{\bibinfo{title}{Introduction to statistical time series}} (\bibinfo{publisher}{John Wiley \& Sons}, \bibinfo{address}{New York}, \bibinfo{year}{2009}).

\bibitem{noether1963note}
\bibinfo{author}{Noether, G.~E.}
\newblock \bibinfo{journal}{\bibinfo{title}{Note on the kolmogorov statistic in the discrete case}}.
\newblock {\emph{\JournalTitle{Metrika}}} \textbf{\bibinfo{volume}{7}}, \bibinfo{pages}{115--116} (\bibinfo{year}{1963}).

\end{thebibliography}

\section*{Data Availability Statement}
Data is collected from \url{https://seklima.met.no/} and no new data were created for this study.

\section*{Acknowledgment}
T.C. was supported by Sorbonne Research Institute Funding. D.G. was supported by the Science and Engineering Research Board (SERB), Government of India (Project No. CRG/2021/005894).	

\section*{Author contributions statement}

The following is a list of individuals and their respective contributions to this manuscript:\\

Conceptualization: Arnob Ray, Tanujit Chakraborty, Athulya Radhakrishnan, Chittaranjan Hens, Syamal K Dana, Dibakar Ghosh, Nuncio Murukesh.

Data curation: Arnob Ray, Tanujit Chakraborty.

Formal analysis: Arnob Ray, Tanujit Chakraborty.

Funding acquisition: Tanujit Chakraborty.

Investigation: Arnob Ray, Tanujit Chakraborty, Chittaranjan Hens.

Methodology: Arnob Ray, Tanujit Chakraborty, Chittaranjan Hens.

Project administration: Syamal K Dana, Dibakar Ghosh, Nuncio Murukesh.

Resources: Arnob Ray, Tanujit Chakraborty, Chittaranjan Hens.

Software: Arnob Ray, Tanujit Chakraborty.

Supervision: Syamal K Dana, Dibakar Ghosh, Nuncio Murukesh.

Validation: Arnob Ray, Tanujit Chakraborty, Chittaranjan.

Visualization: Arnob Ray, Tanujit Chakraborty, Chittaranjan Hens, Syamal K Dana, Dibakar Ghosh, Nuncio Murukesh.

Writing – original draft: Arnob Ray, Tanujit Chakraborty.

Writing – review \& editing: Athulya Radhakrishnan, Chittaranjan Hens, Syamal K Dana, Dibakar Ghosh, Nuncio Murukesh.

	\section*{APPENDIX}\label{sec:appendix}
\par 
{\it Mann-Kendall (MK) test}:
	To determine whether a trend is present in time series data, a Mann-Kendall trend test is applied \cite{gocic2013analysis}. The hypotheses are defined as:\\
	$H_0$ (null hypothesis): there is no trend present in the data and
	$H_1$ (alternative hypothesis): a trend is present in the data. \\
	The presence of a trend in the time series data is statistically supported if the $p$-value of the test is less than a predetermined level of significance such as $\alpha=0.05$ because the null hypothesis is rejected.
 
	{\it Augmented Dickey-Fuller (ADF) Test}: It is a statistical test for determining whether or not a time series is stationary \cite{fuller2009introduction}. The null hypothesis $(H_0)$ indicates that the time series is regarded as non-stationary and the alternate hypothesis ($H_1$) demonstrates that the time series is referred to as stationary. We can reject the null hypothesis and conclude that the time series is stationary if the $p$-value from this test is less than a specific threshold or significance level (say, $\alpha = 0.05$).
	
	{\it Kolmogorov-Smirvov (KS) Goodness-of-Fit Test}: 
	The Kolmogorov-Smirnov test is used to figure out whether or not a sample belongs to a certain distribution \cite{noether1963note}. The hypotheses are as follows: The null hypothesis ($H_0$) implies that the dataset follows a given distribution, whereas the alternative hypothesis ($H_1$) asserts that it does not. If the $p$-value is greater than the level of significance ($\alpha=0.05$), we fail to reject the null hypothesis.






\end{document}